\documentclass[%
 aip,
 amsmath,amssymb,
 reprint,%
]{revtex4-1}

\usepackage{graphicx}
\usepackage{dcolumn}
\usepackage{bm}
\usepackage[normalem]{ulem}
\usepackage[utf8]{inputenc}
\usepackage[T1]{fontenc}
\usepackage{mathptmx}
\usepackage{etoolbox}

\usepackage[colorlinks]{hyperref}
\usepackage[noabbrev]{cleveref}
\usepackage{listings, color}
\usepackage{enumitem}
\usepackage{float}
\usepackage{textcomp}
\usepackage{graphicx}
\usepackage{appendix}
\usepackage{todonotes}
\usepackage{ragged2e}
\graphicspath{{Figures/}}
\newcommand{\COMMENT}[1]{}

\makeatletter
\def\@email#1#2{%
 \endgroup
 \patchcmd{\titleblock@produce}
  {\frontmatter@RRAPformat}
  {\frontmatter@RRAPformat{\produce@RRAP{*#1\href{mailto:#2}{#2}}}\frontmatter@RRAPformat}
  {}{}
}%
\makeatother
\begin{document}

\preprint{AIP/123-QED}

\title[Machine Learning Changes the Rules for Flux Limiters]{Machine Learning Changes the Rules for Flux Limiters}
\author{Nga Nguyen-Fotiadis}
\affiliation{ 
Information Sciences, CCS-3, Los Alamos National Laboratory, Los Alamos, NM 87545, USA
}%

\author{Michael McKerns}%
\affiliation{ 
Information Sciences, CCS-3, Los Alamos National Laboratory, Los Alamos, NM 87545, USA
}%

\author{Andrew Sornborger}
\email{sornborg@lanl.gov}
\affiliation{ 
Information Sciences, CCS-3, Los Alamos National Laboratory, Los Alamos, NM 87545, USA
}%


\begin{abstract}
Learning to integrate non-linear equations from highly resolved direct numerical simulations (DNSs) has seen recent interest for reducing the computational load for fluid simulations. Here, we focus on determining a flux-limiter for shock capturing methods. Focusing on flux limiters provides a specific plug-and-play component for existing numerical methods. Since their introduction, an array of flux limiters has been designed. Using the coarse-grained Burgers' equation, we show that flux-limiters may be rank-ordered in terms of {\color{black} their log-error relative to high-resolution data.}
We then develop theory to find an optimal flux-limiter and present flux-limiters that outperform others tested for integrating Burgers' equation on lattices with $2\times$, $3\times$, $4\times$, and $8\times$ coarse-grainings.
{\color{black} We train a continuous piecewise linear limiter by minimizing the mean-squared misfit to 6-grid point segments of high-resolution data, averaged over all segments.
While flux limiters are generally designed to have an output of $\phi(r) = 1$ at a flux ratio of $r = 1$, our limiters are not bound by this rule, and yet produce a smaller error than standard limiters.
We find that} our machine learned limiters have distinctive features that may provide new rules-of-thumb for the development of improved limiters.
Additionally, we {\color{black} use our theory to learn flux-limiters that outperform standard limiters across a range of values (as opposed to at a specific fixed value) of} coarse-graining, number of discretized bins, and diffusion parameter. {\color{black} This demonstrates the ability to produce flux limiters that should be} more broadly useful {\color{black} than standard limiters} for general applications.
\end{abstract}

\maketitle

\section{Introduction}
Numerically integrating fluid equations on a coarse grid relative to a fully resolved integration of a given problem (coarse-graining) is of both historical and current interest \cite{besnard1992turbulence,schwarzkopf2011application,zhou2022preconditioned,valle2022implementation,abedian2022third} since it can significantly reduce the computational time required for a given simulation.
With the advent of modern machine learning methods, a number of attempts have been made to develop improved coarse-grained models \cite{vinuesa2022enhancing}, including for 3D Eulerian \cite{tompson2017accelerating} and Navier-Stokes turbulence \cite{mohan2019compressed}, buoyancy-driven, variable density turbulence \cite{nadiga2019leveraging}, and molecular-level simulations \cite{chan2019machine}.

In this paper, our focus will be machine learning an accurate local integrator for the coarse-grained, 1-D viscous Burgers' equation  \cite{burgers1948mathematical}. Burgers' equation is the simplest fluid equation admitting shocks \cite{eberhard1942partial,cole1951quasi} and is exactly integrable. 
Numerically integrating shocks is complicated by the Gibbs effect \cite{wilbraham1848certain}, where, when a discontinuity develops, unstable oscillations develop in a numerical simulation.
This behavior may be corrected using shock capturing methods \cite{godunov1959difference,van1979towards,colella1984piecewise,harten1997high,shu1988efficient}. 
Shock capturing methods rely on {\it flux limiters} - non-linear interpolations between high- and low-resolution integration schemes used in a numerical simulation to keep a shock solution monotonic, thereby eliminating spurious oscillations.

Harten \cite{harten1997high} provided a framework for constructing non-linear, monotonicity preserving flux limiters.
At present, a large number of flux limiters have been defined and used in the literature \cite{zhang2015review}.
Different forms of flux limiter have been shown to have differing accuracy and convergence performance \cite{zhang2015review}. 
And criteria have been studied for determining what parameter regions are appropriate for different classes of limiter \cite{sweby1984high}.

Since Burgers' equation is exactly integrable, it is straightforward to generate accurate, high-resolution data from which flux limiters may be tested using a machine learning approach. 
This setting also provides a good testing ground for developing methodology and evaluating which flux limiters are the best (for one particular equation, or potentially for multiple shock-forming equations), and for comparing existing flux limiters with those learned from data.

Below, we will introduce the Cole-Hopf solution to the inviscid Burgers' equation, show how to use it to test DNS training data, and define our flux limiters.
We will then present a cost function allowing us to compare predicted outcomes from a discretized flux limiter with high-resolution DNS training data.
Next, we will optimize the cost function for a given set of data resulting in a regression analysis solvable with standard linear algebraic methods.
We then apply this analysis to a training dataset to learn optimal flux limiters from coarse-grained data.
We compare these optimized, coarse-grained flux limiters with a large set of other flux limiters and show that they outperform them over the set of coarse-grainings.
We finally incorporate our regression analysis within a larger, hyper-parameter optimization framework to find an optimal discretization of the flux limiter over a range of parameters, including coarse-graining, number of discretization bins, and diffusion parameter.

\begin{figure*}[ht]%
    \includegraphics[width=1.0\textwidth]{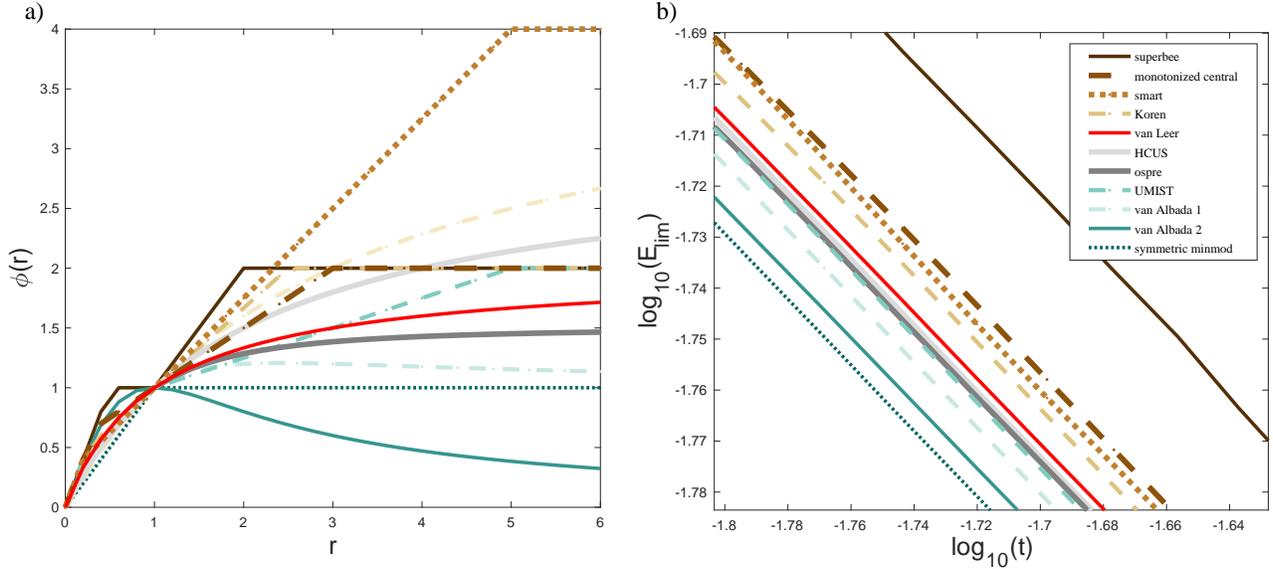}
    \caption[justification=left]{ A set of flux limiters and their log-error relative to high resolution data. In this figure and in Fig.~\ref{fig:across_bin2x}, we use the same colors and patterns to indicate a particular flux limiter. For instance, the van Leer limiter is plotted in solid red. The legend for both panel a) and panel b) is shown in panel b). a): 11 flux limiters, $\phi_i(r)$, where $i$ ranges over a set of standard flux limiters (see legend in b)), plotted as a function of flux ratio $r$. All flux limiters, by design \cite{sweby1984high}, go through the point, $(r,\phi) = (1,1)$.  b): Log-error computed for each standard limiter in a) with error, $E_\mathrm{lim} = e_i - e_\mathrm{hres}$, with $e_i$ the shock capturing solution with limiter, $i$, and $e_\mathrm{hres}$ the $2\times$ coarse-grained high-resolution solution. Here, we compare against the standard sinusoidal solution of Burgers' equation. Note that while most limiters are bowed out below $r < 1$, the better performing (lower error, in green) limiters tend to be comparatively (with regard to van Leer, in red) bent down for $r > 1$.}
    \label{fig:FL}
\end{figure*}
\section{Background}
\subsection{Exact Solution of Inviscid Burgers' Equations}
The viscous Burgers' equation \cite{burgers1948mathematical} written in conservative form is
\begin{equation}
  \frac{\partial u}{\partial t} + \frac{\partial}{\partial x}\left( \frac{u^2}{2} - \nu \frac{\partial u}{\partial x}\right) = 0 \; .
\end{equation}
It may be solved exactly with the Cole-Hopf transform \cite{cole1951quasi},
\begin{equation}
  u = -2\nu \frac{1}{\phi}\frac{\partial \phi}{\partial x} \; .
\end{equation}
For periodic boundary conditions (those considered here), the transform results in a diffusion equation
\begin{equation}
  \frac{\partial \phi}{\partial t} = \nu \frac{\partial^2 \phi}{\partial x^2} \; ,
\end{equation}
which may be solved with Fourier methods. 
Finally, the Cole-Hopf transform is inverted to recover the solution in the original coordinates.

We used a discretized version of the Cole-Hopf transform on a finely resolved lattice to generate simulation data. This approach works well for moderately sized lattices, but has instabilities for larger lattices when $\nu$ becomes small \cite{maritz1991exact}. Thus, for our flux limiter optimization procedure, below, we used a first-order, but very accurate scheme to generate our training and test data and validated the scheme, which is stable for any sized lattice, on simulations on smaller lattices by comparing with the discretized Cole-Hopf transform (See Appendix~\ref{apdx:first_and_exact}).
We call data generated in this way {\it exact} or {\it high-resolution} data.
Note that this solution method can resolve shocks without the need for flux limiters and thus provides good data to test flux limiters in shock capturing integration schemes. For equations with no analytical solution, high-resolution data from finely resolved DNSs would need to be used.

\subsection{Flux Limiters} \label{sec:FluxLimiter}
We use a semi-discrete scheme to integrate initial conditions from coarse-grained data selected at different resolutions from previously generated exact data. The integration scheme takes the general form
\begin{equation}
  \frac{du_i}{dt} + \frac{1}{\Delta x_i} \left[ G\left(u_{i+\frac{1}{2}}\right) - G\left(u_{i-\frac{1}{2}}\right) \right]  = 0\; .
  \label{eq:burgers}
\end{equation}
Here, $i$ is a cell index, and $G\left(u_{i-\frac{1}{2}}\right)$ and $G\left(u_{i+\frac{1}{2}}\right)$ denote edge fluxes.
The edge fluxes are non-linear interpolations between high- and low-resolution fluxes
\begin{eqnarray}
    G\left(u_{i+\frac{1}{2}}\right) & = & \left(1 - \phi(r_{i})\right) f^\mathrm{low}_{i+\frac{1}{2}}  + \phi(r_{i}) f^\mathrm{high}_{i+\frac{1}{2}},  \\
    G\left(u_{i-\frac{1}{2}}\right) & = & \left(1 - \phi(r_{i-1})\right) f^\mathrm{low}_{i-\frac{1}{2}}  + \phi(r_{i-1}) f^\mathrm{high}_{i-\frac{1}{2}}\; .
\end{eqnarray}
Here, $f^\mathrm{low}$ indicates a low-resolution flux, typically first-order, which does not suffer from Gibbs phenomena, and $f^\mathrm{high}$ indicates a high-resolution flux to integrate smoother regions of a solution.
The flux limiter, $\phi(r)$, is a function of the flux ratio, 
\begin{equation}
  r_i = \frac{u_i - u_{i-1}}{u_{i+1} - u_i} \; .
    \label{eq:ratio}
\end{equation}
A large number of flux limiters have been proposed and used. For instance, the van Leer flux limiter \cite{van1974towards} takes the form
\begin{equation}
  \phi(r) = \frac{r + |r|}{1 + |r|} \; .
\end{equation}
Note that this flux limiter and all of the flux limiters that we consider (see Fig.~\ref{fig:FL} and Tab.~\ref{tab:fl_table}) here are (piecewise) continuous and zero for $r \le 0$.

\section{Methods} 
\subsection{A Shock Capturing Integration Method for Burgers' Equation}
For low- and high-resolution, we chose Lax–Friedrichs (LF)
\begin{eqnarray}
    f^\mathrm{low}_{i \pm \frac{1}{2}} = f^{\mathrm{LF}}_{i \pm \frac{1}{2}} =&& \frac{1}{2} [ F(u_i) + F(u_{i+1}) \mp \alpha \frac{\Delta x}{\Delta t}(u_{i\pm 1} -u_i) ]; \nonumber \\ &&\alpha  = \max_{{u}} |\frac{\partial F}{\partial {u}}|  
    \label{eq:Lax-F}
\end{eqnarray}
and Lax-Wendroff (LW) fluxes
\begin{eqnarray}
    f^\mathrm{high}_{i \pm \frac{1}{2}} = f^{\mathrm{LW}}_{i \pm \frac{1}{2}} &&= \frac{1}{2} [F(u_{i}) + F(u_{i+1}) \nonumber \\ 
            &&\mp \frac{\Delta t}{\Delta x} \left(\frac{\partial F}{\partial u} (u_{i \pm \frac{1}{2}})\right) ( F(u_{i \pm 1}) - F(u_{i}))  ],
    \label{eq:Lax-W}
\end{eqnarray} 
where $F = \frac{u^2}{2} - \nu \frac{\partial{u}}{\partial{x}}$ is the flux from Burgers' equation.  Eq.~\eqref{eq:burgers} now becomes:
\begin{equation}
    u_i(t_{n+1}) = u_i(t_n)  - \frac{\Delta{t}}{\Delta x} {\Delta{F}}^i
\end{equation}
with
\begin{eqnarray}
    \label{eq:DeltaF}
    {\Delta F}^i &=& {\Delta{F}_1}^i + \phi(r_i){\Delta{F}_2}^i + \phi(r_{i-1}){\Delta{F}_3}^i \; ,
\end{eqnarray}
where
${\Delta{F}_1}^i$, ${\Delta{F}_2}^i$, and ${\Delta{F}_3}^i$ can be written explicitly as:
\begin{eqnarray}
    {\Delta{F}_1}^i &=& f^{\mathrm{LF}}_{i+\frac{1}{2}} - f^{\mathrm{LF}}_{i-\frac{1}{2}}, \nonumber \\ 
    {\Delta{F}_2}^i &=&f^{\mathrm{LW}}_{i+\frac{1}{2}} - f^{\mathrm{LF}}_{i+\frac{1}{2}},  \nonumber \\
    {\Delta{F}_3}^i &=& -(f^{\mathrm{LW}}_{i-\frac{1}{2}} - f^{\mathrm{LF}}_{i-\frac{1}{2}}). 
    \label{eq:3DeltaF}
\end{eqnarray}

In Fig.~\ref{fig:FL}(a), we plot a set of $12$ different flux-limiters from the literature. An important feature to note in these flux limiters is that they all pass through the point $\phi(1) = 1$. This is due to a requirement of second-order accuracy of the shock capturing scheme and Lipshitz continuity of $\phi(r)$ \cite{sweby1984high}. In Fig.~\ref{fig:FL}(b), we show the difference between solutions computed with the various flux limiters and high-resolution data. In this example, both simulation and high-resolution data are coarse grained at $2 \times$ (i.e. initial conditions for each simulation were subsampled at every other grid point from high-resolution data, integrated with a shock-capturing scheme for time $t$, then compared to subsampled high-resolution data at time $t$). In (Fig.~\ref{fig:FL}(b)), we see an ordering where some flux limiters perform better than others for predicting coarse-grained data. Note that, although we plot the errors over a short time, the errors retain the same ordering over the whole integration. Thus, we can define an ordering of the limiters.
%
%

At this point, the question as to what the optimal flux limiter is arises. In this study, we set up the problem of how to find a flux limiter, $\phi (r)$, that is optimal  with respect to certain criteria.  We will train a piecewise linear flux limiter using an exact dataset (Appendix \ref{apdx:first_and_exact}) and compare it with the $11$ flux limiters shown in Fig.~\ref{fig:FL}(a).

\begin{figure*}[ht]
    \includegraphics[width=0.80\textwidth]{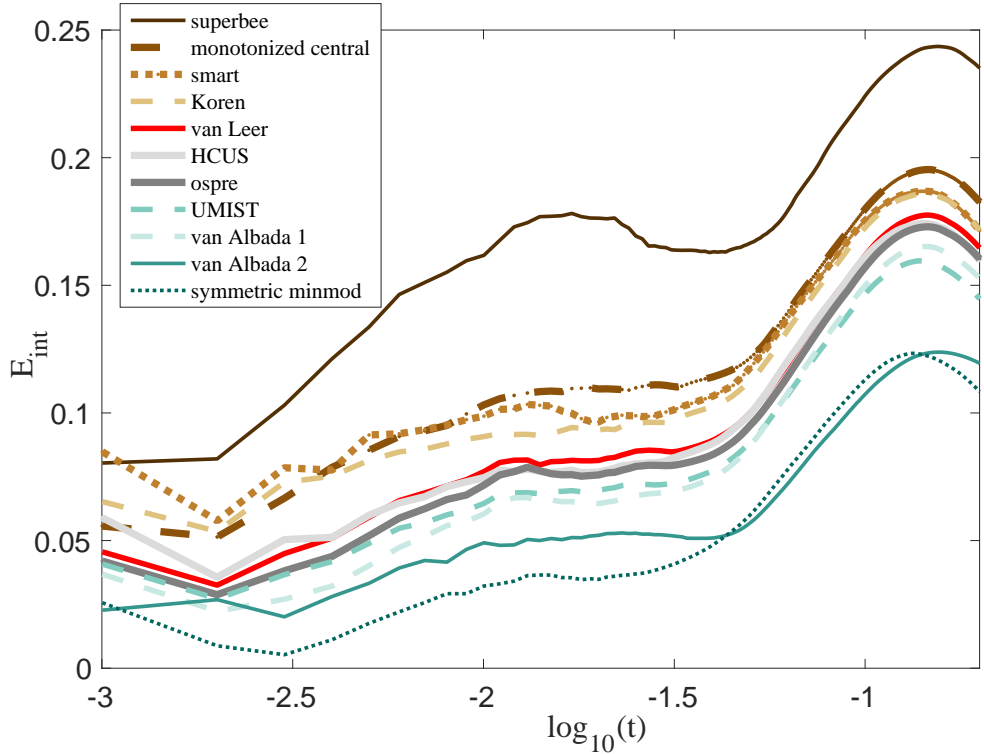}
    \caption{Learning a $2 \times$ coarse-grained flux limiter. 
    We plot the error relative to our machine learned flux limiter (so, our flux limiter's error relative to itself is zero, indicated by the $x$-axis) integrated across the spatial dimension, $E_\mathrm{int}$, calculated for a set of standard flux limiters using $K=20$ bins versus (log of) time. Here, a positive value for $E_\mathrm{int}$ indicates worse accuracy for the limiter being compared to. This means that, compared to all standard limiters, the $2 \times$, $K = 20$ bin learned limiter performs better. Also, note that, similarly to what we saw in Fig.~\ref{fig:FL}, the standard limiters are ordered with respect to their performance with some performing better and some worse relative to the machine learned limiter.
    }
    \label{fig:across_bin2x}
    \vspace{-0.1 in}
\end{figure*}
\subsection{Flux Limiter Discretization}

We discretize the flux-limiter that we will optimize with machine learning methods, $\phi (r)$, in piecewise linear segments, where the $k$'th segment has the form, 
\begin{eqnarray}
    \phi_k (r)  &&= \phi_0 + b_1 (r_2 - r_1) + b_2 (r_3 - r_2) + ... + b_k (r - r_k) \nonumber \\ 
                &&+ 0_{k+1} + ... +0_{K},
    \label{eq:pw-linear}
\end{eqnarray}
and $r \in [r_k, r_{k+1})$, $k \in \{1, \dots, K\}$, $\phi_0 = 0$, and $b_i$ are slope coefficients. Note that for $r \le 0$, $\phi(r) = 0$ and for $r = r_K$, all terms in Eq.~\eqref{eq:pw-linear} are non-zero. Below, we use vector notation, $\boldsymbol{b} = [b_1, b_2, ..., b_k, b_{k+1}, ..., b_K]^{\mathrm{T}}$ for slope coefficients.  Eq.~\eqref{eq:pw-linear} can be rewritten as $ \phi_k(r) = \boldsymbol{b}^{\mathrm{T}} \boldsymbol{\Delta{r}}_k $
with $\boldsymbol{\Delta{r}}_k$ defined as
\begin{equation}
\boldsymbol{\Delta{r}}_k = [r_2 - r_1, r_3 - r_2, ..., r - r_k, 0,..., 0]^{\mathrm{T}}.
\label{eq:Deltar}
\end{equation}

\subsection{Learning an Optimal Discretized Flux Limiter} \label{subsec:costfunc}
To optimize the discretized flux-limiter in Eq.~\eqref{eq:pw-linear}, we define the mean squared error between $N$ input-output pairs, $\{ o_i(\{ u_{c}^i \}), g_i \}$:
\begin{equation}
    C = \frac{1}{2} \sum_{i=1}^{N} { ( o_i(\{ u_{c}^i \}) - g_i  )^2 }
    \label{eq:loss}
\end{equation}
as the cost. Here, $g_i$ is the high-resolution fluid velocity at the $i$-th grid position at time $t_{n+1}$ and $o_i$ is the shock-capturing method's prediction of the fluid velocity at time $t_{n+1}$ from data at the previous timestep. $o_i$ is a functional of a subset of data points $\{ u_c^i \} = \{u_{c}^{i1}, u_{c}^{i2}, u_{c}^{i3}, ..., u_{c}^{iN_c} \}$ indicated relative to the $i$-th grid position at time step $t_n$.  Here, we used $N_c = 6$ data points at time $t_n$ (see details in Appendix \ref{apdx:customized_FL}) to predict a data point $g_i$ at $t_{n+1}$, i.e. $ \{ u_c^i \} = \{ u_{i-3}, u_{i-2}, u_{i-1}, u_{i}, u_{i+1}, u_{i+2} \}$.  Thus, $o_i(\{ u_c^i\})$ is the integration obtained with the flux-limiter method defined in Eqs.~(\ref{eq:burgers}), ~(\ref{eq:Lax-F}), ~(\ref{eq:Lax-W}), ~(\ref{eq:pw-linear}) given a set of 6-points $\{u_c^i \}$:
\begin{equation}
    o_i (\{ u_c^i\}, t_{n+1}) = u_i(t_n) - \frac{\Delta t}{\Delta x} \Delta F (\{ u_c^i \}, \{b_i \}, t_n).
    \label{eq:o_i}
\end{equation}
Here, $\Delta F (\{ u_c^i \}, \{b_i \}, t_n)$, defined via Eqs.~\eqref{eq:DeltaF} and \eqref{eq:3DeltaF}, 
is the difference of the two fluxes defined in Eq.~\eqref{eq:burgers}.  The minimum of the cost function, Eq.~\eqref{eq:loss}, can be computed exactly by finding the unique root, $\boldsymbol{b}$, of the equation $ \frac{\partial L}{\partial {\boldsymbol{b}}} = \boldsymbol{0} $, that is:
\begin{equation}
    \sum_{i=1}^{N} \left(u_i - g_i - \frac{\Delta{t}}{\Delta{x}} \Delta{F}^i\right) \left(-\frac{\Delta{t}}{\Delta{x}}\right) \boldsymbol{\Delta{s}}_i \boldsymbol{\Delta{F}}^{i}_{2,3} = \boldsymbol{0}.
    \label{eq:root}
\end{equation}
In Eq.~\eqref{eq:root}, $\Delta{F}^i = \Delta F (\{ u_c^i \}, \{b_i \}, t_n)$ is defined via Eqs.~\eqref{eq:DeltaF} and ~\eqref{eq:3DeltaF}.  $\boldsymbol{\Delta{s}}_i = [\boldsymbol{\Delta{r}}_i, \boldsymbol{\Delta{r}}_{i-1} ]$ is a $K \times 2$ matrix with $\boldsymbol{\Delta{r}}_i$ defined in Eq.~\eqref{eq:Deltar}. $\boldsymbol{\Delta{F}}_{2,3}^i = [ \Delta{F}^i_2, \Delta{F}_3^i ]^{\mathrm{T}}$ with components $\Delta{F}_{2}^i$ and $\Delta{F}_{3}^i$ defined via Eq.~\eqref{eq:3DeltaF}.

\begin{figure*}[ht]%
    \includegraphics[width=1.0\textwidth]{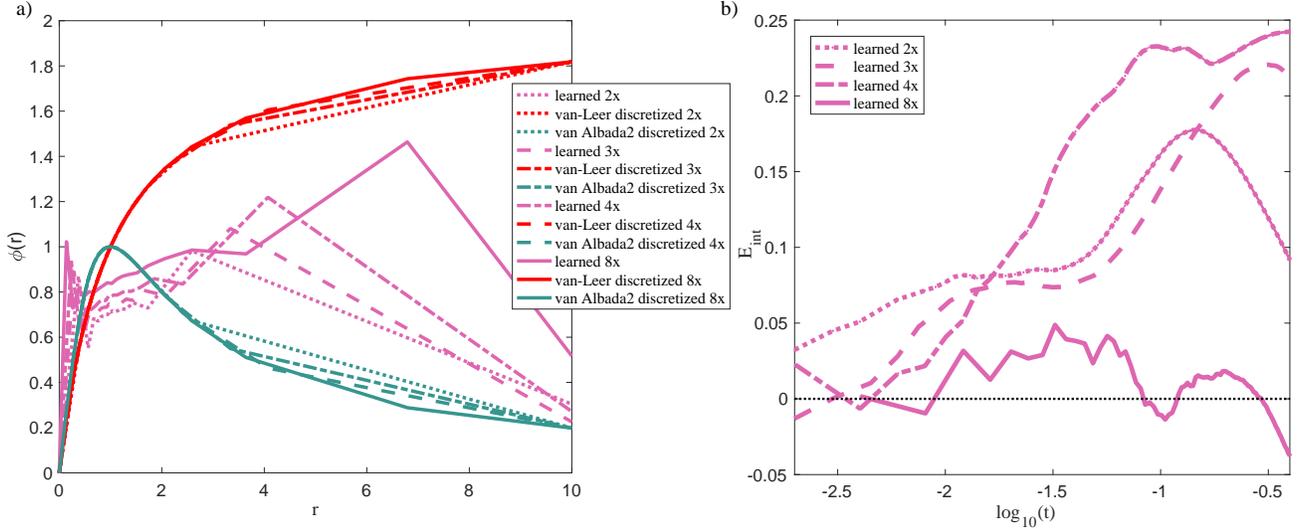}
    \caption{(a): $K = 20$ bin machine learned limiter functions $\phi (\boldsymbol{r})$, van Leer, and van Albada 2 limiters plotted using the same discretized bins from $4$ different coarse grained cases: $2 \times$, $3 \times$, $4 \times$, and $8 \times$. (b): Corresponding integrated relative errors, $E_\mathrm{int}$ (\ref{relerrvL}), between the learned flux limiters  with van Leer as a function of (log) time as compared to the ground truth for $2 \times$, $3 \times$, $4\times$, and $8\times$. Note that the machine learned limiters are not constrained to go though $(r,\phi) = (1,1)$ (and, in this case, they do not).}
    \label{fig:2x3x4x8x_FLfuncs_and_errors}%
\end{figure*}
Solving Eq.~\eqref{eq:root} reduces to solving a linear equation $ \boldsymbol{A} \cdot \boldsymbol{b} = \boldsymbol{C} $ that yields $\boldsymbol{b} = \boldsymbol{A}^{-1} \cdot \boldsymbol{C}$. Here, $\boldsymbol{A} = \boldsymbol{\Delta{r_F}} \cdot (\boldsymbol{\Delta{r_F}})^{\mathrm{T}} $ and $\boldsymbol{C} = \frac{\Delta{x}}{\Delta{t}}\sum_{i=1}^{N} { {{O_G}^i} \boldsymbol{\Delta{r_F}}^i }$, where  $\boldsymbol{\Delta{r_F}}$ is a $K \times N$ matrix with each column $\boldsymbol{\Delta{r_F}}^i$ a $K \times 1$ vector defined as $\boldsymbol{\Delta{r_F}}^i = (\boldsymbol{\Delta{s}}_i) (\boldsymbol{\Delta{F}}_{2,3}^i)$. Finally, ${O_G}^i = u_i - g_i - \frac{\Delta{t}}{\Delta{x}} \Delta{F}_1^i$.  Note that $\Delta{F}_1^i$ is defined via Eq.~\eqref{eq:3DeltaF} and we recall that $K$ is the size of the discretized flux limiter (i.e. the size of $\boldsymbol {b}$).  Hence, each matrix $\boldsymbol A$ (or $\boldsymbol C$) is a function of $N$ training data points. 

We wish our estimates of each segment of the flux limiter, $\phi_k(r)$, to have the same variance. Choosing this discretized space wisely is an important step.  Our choice was to discretize the flux limiter such that each segment contained an equal number of training data points.  

We used the above method with different coarse-grainings of the high-resolution dataset as training data in order to find optimal flux limiters, $\phi_k(r)$, for a set of coarse-grainings.

\subsection{Hyperparameter Optimization of a Discretized Flux Limiter} \label{hyperopt}

The cost function defined in the previous section is intended to optimize
a discretized flux-limiter for a given coarse-graining, $CG$, number of bins, $K$,
and diffusion parameter, $\mu$. We automated the generation of training and test
data, the training of a discretized flux-limiter, and the testing of
the learned limiter, to yield a function that produces the cost $C$ as defined in Eq. \eqref{eq:loss} for a given $CG$, $K$, and $\mu$.
This approach leads to very good flux limiters for a particular set of parameters, however, the question remains as to whether we can find limiters that function well in a more general context. To attack this issue, we extended  our approach by optimizing over (hyper)parameters. By averaging the cost for a learned flux-limiter over all segments, 
\begin{equation}
    \overline{C} = \sum_{k=1}^{K} { C_{k} / K } \; .
    \label{eq:aveloss}
\end{equation}
In particular, we used an optimizer to produce a flux-limiter that
minimizes $\overline{C}$ within the region defined by
{\color{black}$\mu \in [0.005, 0.0248]$}, $CG \in [2, 10]$, and {\color{black}$K \in [2, 38]$}.


\begin{figure*}[ht]%
    \includegraphics[width=0.7\textwidth]{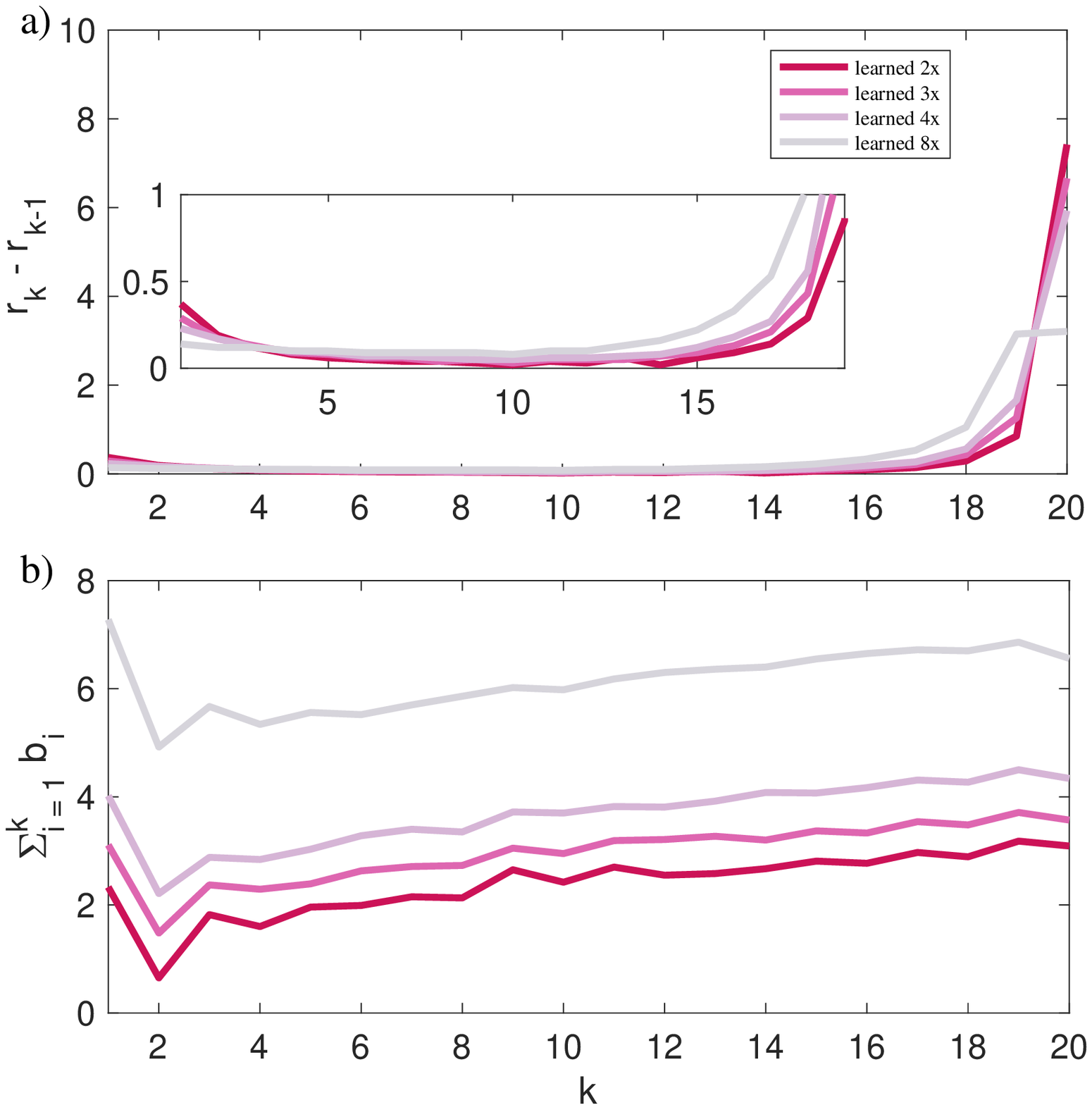}
    \caption{Patterns in the scaling of learned flux limiters. a) Bin widths vs. $k$. Note the inversion of bin widths from left to right. Larger coarse grainings have smaller bin widths close to $r = 0$ and vice-versa for large $r$. b) The cumulative sum of the slopes, $\sum_{i = 1}^k b_i$. This function has the same form across coarse-grainings but scales as a function of coarse-graining. Note also that $b_1$ is proportional to the coarse-graining, and the bulk of the $k$ points occur in the roughly linear region with low-amplitude fluctuations ($r \in [0.5,2.0]$) in Fig.~\ref{fig:2x3x4x8x_FLfuncs_and_errors}(a).}%
    \label{pattern}
\end{figure*}

\section{Results}
We generated $500$ Burgers' simulations with $500$ random initial conditions that have in total $160$M data points. We attained good convergence of training results even with just $80$ simulations (i.e. using less than $30$M data points).  For a discretization into $20$ segments, the solution to $\boldsymbol b$ was estimated with $1.5$M data points per element and hence had a standard error of $0.0008$. For this case, diagonalization was performed on the inverse of the square matrix $\boldsymbol A$ of size $K \times K$ with maximal $K$ being $K=20$ segments.  Better standard errors were obtained with fewer segments, but at the cost of worse resolution of the limiter. 

We validated our learning model on a subset of hold-out data which contained samples, $\{ u_c^i \}$, from about $20$ simulations. 

We compared our optimized flux limiter with a set of standard flux limiters (sFLs). We computed the spatially integrated relative error
\begin{equation}
  E_\mathrm{int} \equiv \sum_i \left(\frac{e_i^\mathrm{sFL} - e_i^\mathrm{learned} }{e_i^\mathrm{learned}} \right) \; ,
  \label{relerrvL}
\end{equation}
where the error at a given grid location, $i$, for each flux limiter is
\begin{equation}
  e_i = o_i - g_i \; .
  \label{eq:error}
\end{equation}


Note that this error, as defined, is positive when the learned limiter outperforms a given sFL. We investigated $K = 2$, $5$, and $20$ segment flux limiters. Optimized flux limiters for all values of $K$ were better than the discretized van Leer limiter (i.e. the relative error was everywhere $> 0$ with an average improvement of about $10\%$). 


In Fig.~\ref{fig:across_bin2x}, we plot the integrated relative error of our optimized flux limiter with $K = 20$ versus a set of eleven sFLs. Here, $CG$ is $2 \times$ and $r \in [0.0, 10.0]$. Note that the optimized limiter performs better than all other limiters investigated. Here, the symmetric minmod limiter performs best, but still reaches $10\%$ greater error relative to the optimized limiter.


We plot optimized limiters, $\phi_k(r)$ in Fig.~\ref{fig:2x3x4x8x_FLfuncs_and_errors}(a). 
\begin{figure*}[ht]%
    \includegraphics[width=0.95\textwidth]{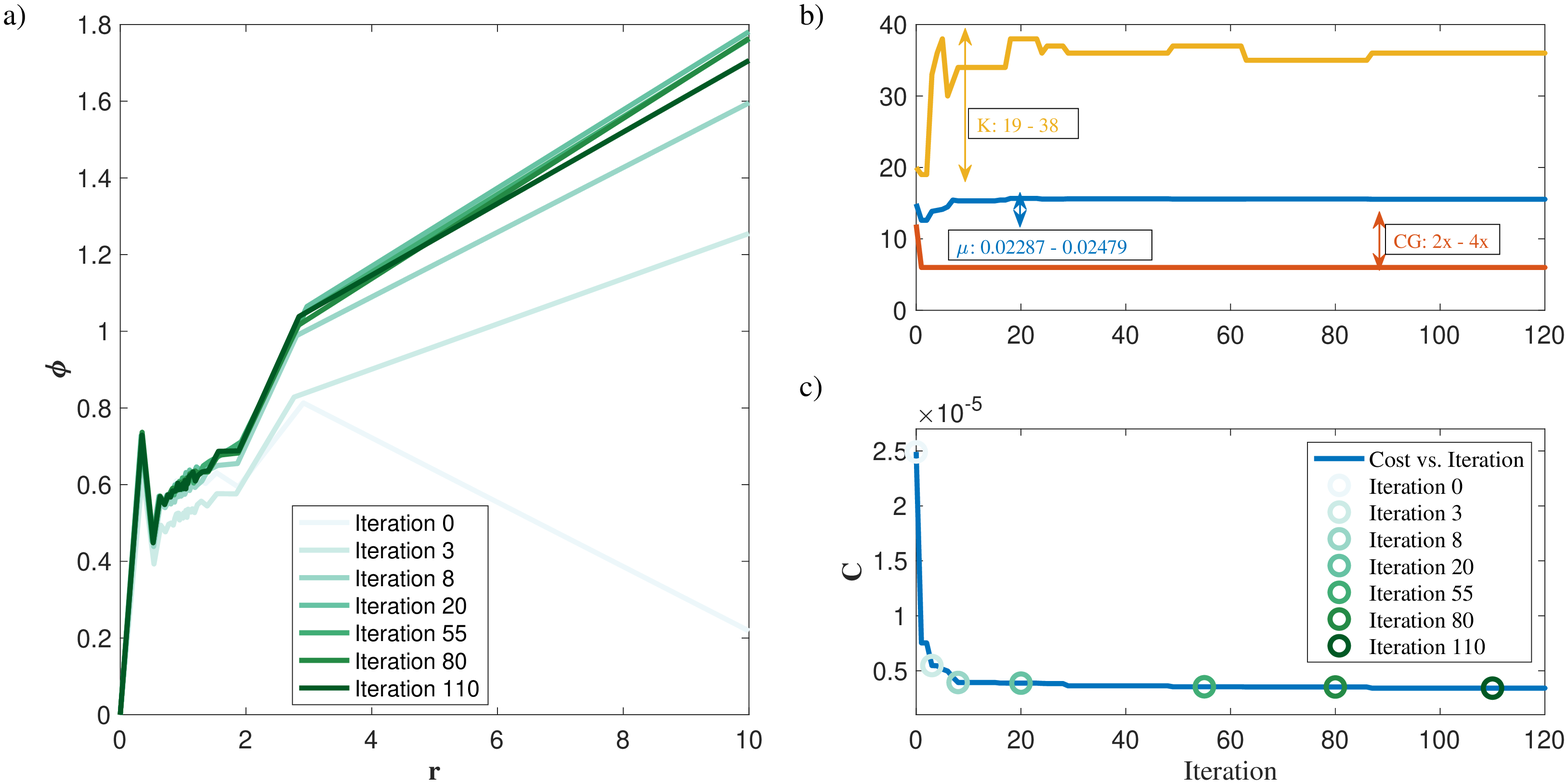}
    \caption{Finding optimal flux limiters across multiple hyperparameters. a) A sequence of flux limiters learned over multiple iterations of optimization. The legend indicates the coarse-graining and the number of bins for which the flux limiter performs best, and the iteration at which it was found. b) Parameter convergence vs. iteration. Double headed arrows indicate the scale of parameter variation. Number of bins (yellow), coarse graining (red), diffusion parameter, $\mu$ (blue). c) The cost as a function of iteration. Circles denote iteration steps for flux limiters plotted in a). {\color{black}Note, in comparing b) and c), that convergence of the cost is primarily driven by }{\color{black}$CG$}{\color{black}, while $K$ appears to have }{\color{black}little}{\color{black} ~effect.}}%
    \label{fig:hyperparam}
\end{figure*}
Also, included in the plot are van Leer and van Albada 2 limiters for reference. These two limiters are commonly-used limiters and bound the learned $K=20$ limiters. For this analysis, equal-variance bins (see Sec.~\ref{subsec:costfunc}) were computed individually for coarse grainings of $2 \times$, $3 \times$, $4 \times$, and $8 \times$ (transitioning from dotted to solid magenta lines in Fig.~\ref{fig:2x3x4x8x_FLfuncs_and_errors}(a)). Note that the smallest coarse grained case ($2 \times$) has the smallest first slope $b_1$ in the first linear piece of $\phi (\{b_i\}, r)$, see Table~\ref{tab:firstslopeFL}.
\begin{table}[h]
\begin{tabular}{|c|c|c|c|c|}
\hline
Coarse graining & 2$\times$   & 3$\times$   & 4$\times$   & 8$\times$   \\ \hline
First slope     & 2.33 & 3.11 & 4.02 & 7.28 \\ \hline
\end{tabular}
\caption{ First slope $b_1$ of the flux limiter obtained for 4 different coarse grainings shown as 4 solid lines in Fig.~\ref{fig:2x3x4x8x_FLfuncs_and_errors}(a). Note that $b_1$ approximates the coarse-graining. }
\label{tab:firstslopeFL}
\end{table}
As we increase the coarse graining to a value larger than $2 \times$, we see an increase in the slope of this first segment where $b_1$ roughly tracks the value of the coarse-graining.

In Fig.~\ref{fig:2x3x4x8x_FLfuncs_and_errors}(b), we show errors, $E_\mathrm{int}$, relative to the van Leer limiter, as a function of time, $t$. Note that for $2 \times$ through $8 \times$ coarse grainings, the optimized limiters perform best, but for $10 \times$ (not plotted in (a)), the discretized van Leer limiter performs better. Note that here, we compare with the continuous van Leer limiter.

Upon inspection of Fig.~\ref{fig:2x3x4x8x_FLfuncs_and_errors}, the general form of the learned limiters is 1) an initial (small $r$) sharply sloped kink that gets sharper as coarse-graining increases, 2) a subsequent roughly linear region with low-amplitude fluctuations, and 3) a final (large $r$) kink that is less sharply sloped than the initial kink. In Fig.~\ref{pattern}, we visualize this pattern by plotting the bin widths (Fig.~\ref{pattern}(a)) and integrated slopes (Fig.~\ref{pattern}(b)), $\sum_{i=1}^K b_i$. The main patterns to point out here are that bin widths decrease for small $r$ and increase for large $r$, while the integrated slopes are virtually identical as a function of coarse-graining, with the exception that they scale roughly with the coarse-graining.
Note also (in Tables~\ref{tab:20rs_2348x} and \ref{tab:20coeffs_2348x}) that $r_k \sim 1.0$ when $k \sim K/2$, $r_K \sim CG$, and $b_1 \sim CG$.

Fig.~\ref{fig:hyperparam} demonstrates the outcome of a flux limiter optimization in the hyperparameter subspace, $\{ CG, K, \mu\}$, defined in Sec.~\ref{hyperopt}. For this optimization, we selected the Differential Evolution \cite{StornPrice:1997} optimizer from
\emph{mystic} \cite{MHA:2009, MSSFA:2011} with the coarse-graining and number of bins
restricted to be integers.
{ \color{black} A population of twenty-two initial points $(CG,~K,~\mu)$, chosen at random
from within the set of valid solutions, was mutated at each iteration. Thus,  
each new generation contained twenty-two candidate solutions per iteration. }
The optimizer continued to generate new candidate solutions until
the change in candidate solutions was less than $10^{-10}$ over $100$ iterations.
{\color{black}We used $50$ simulations to build each training set, and $10$ simulations for each test set, where each simulation was run for $t = 800$ steps.}
New training (and test) data is generated for each new combination of
$(CG,~K,~\mu)$ as part of the automated procedure. 

In Fig.~\ref{fig:hyperparam}(a), we plot the learned flux limiters at select iterations as the hyperparameters converge. {\color{black}Note that as the learned limiters improve (with regard to the cost, in Fig.~\ref{fig:hyperparam}(c)), there is a convergence onto a positive linear slope at large $r$, and a convergence of the mid-segment $r_{K/2}$ of the limiter to roughly $(r, \phi) = (1, {\color{black}0.6})$.
In comparing Fig.~\ref{fig:hyperparam}(b) and Fig.~\ref{fig:hyperparam}(c), it is apparent that the convergence of the cost is primarily driven by the convergence of the }{\color{black}coarse-graining, $CG$}{\color{black}, and secondarily the }{\color{black}diffusion parameter, $\mu$}{\color{black}, while the number of bins, $K$, has little effect} (also see Table~\ref{tab:hyperparam}).
{\color{black}The lack of dependence of the cost on $K$ makes sense, given that for all
learned discretized flux limiters, the bulk of the line segments, $r_i$, occur in the roughly linear region, $r \in [0.5, 2.0]$, where the slope $b_i$ exhibits only low-amplitude fluctuations. Essentially, as long as $K$ is large enough to produce the initial and final kinked segments, any remaining extraneous segments appear to condense in the roughly linear region around $r = 1$.

It also appears that }{\color{black}later in the convergence of the cost, }{\color{black}where $CG$ is fixed and $\mu$ is still converging, that the value of $\mu$ may }{\color{black}primarily impact the }{\color{black}shape of the limiter through $\phi_K$ and the slope $b_K$ for the last line segment.  In early iterations, where $CG$ and $\mu$ are not nearly converged, we can see that at the mid-segment $r_{K/2} = 1.0$, but $\phi_{K/2} \ne {\color{black}0.6}$. As $CG$ and then $\mu$ converge, we find $(r, \phi) = (1, {\color{black}0.6})$.

The flux }{\color{black}limiter that is learned in }{\color{black}the hyperparameter optimization }{\color{black}is }{\color{black}given in Tables~\ref{tab:hyperrs} and \ref{tab:hypercoeffs}, and }{\color{black}roughly follows }{\color{black}the `rules' established for $r$ and $b$ in Tables~\ref{tab:20rs_2348x} and \ref{tab:20coeffs_2348x}.
Specifically, learned limiters have an initial kinked region with a large slope, a final kinked region with a smaller slope, and a smaller roughly linear region around $r_k \sim 1.0$ when $k \sim K/2$. Additionally, we see that generally $r_K \sim CG$ and $b_1 \sim CG$}{\color{black}.}

\section{Discussion}

Here, we presented a framework for the data-driven determination of optimal flux limiters for the coarse-grained Burgers' equation. The framework consists of an internal optimizer computable with numerical linear algebraic methods, which can optionally be further optimized via a hyperparameter search over problem-specific parameters such as the number of bins in the limiter, the coarse-graining of the data, and the diffusion parameter (for Burgers' equation this is the only physical parameter in the problem).

Similar, data-driven approaches to the integration of partial differential equations have been undertaken \cite{bar2019learning}. Here, by focusing on optimizing flux limiters, we attacked a central component of shock-capturing methods for integrating fluids. This allows our flux limiters to be essentially plug-and-play components for many existing numerical codes. We caution, however, that further work must still be done to understand how generally valuable our results are for fluid equations other than Burgers'. Information as to whether and how the regularities discovered in learned limiters change dependent on the target system may be useful for understanding the structure of shock-capturing codes more generally.

\begin{figure*}[ht]%
    \includegraphics[width=1.0\textwidth]{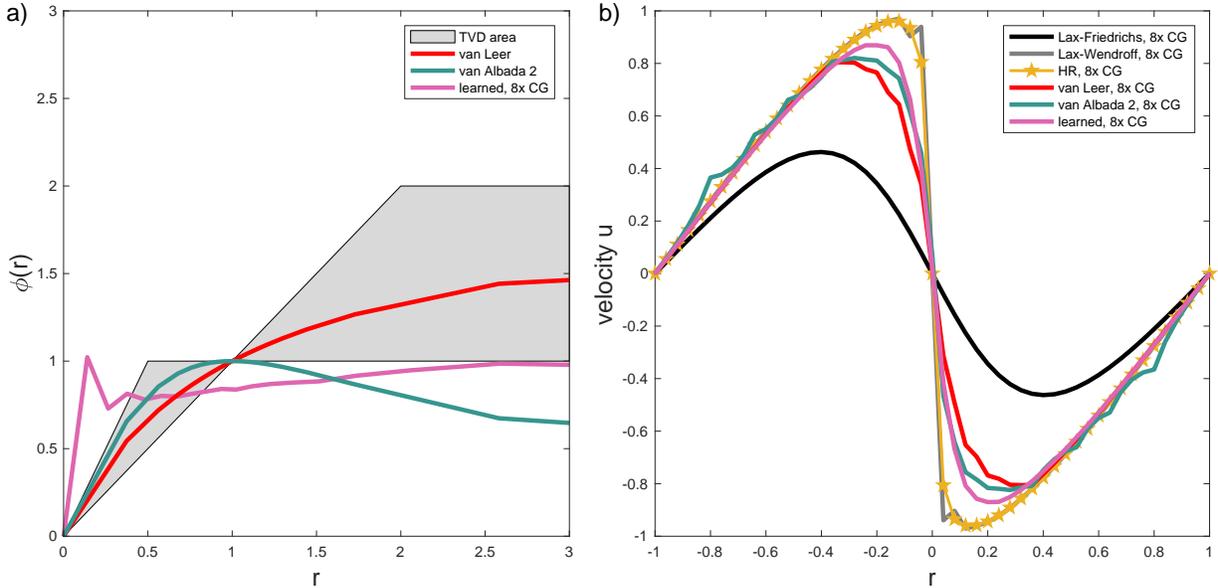}
    \caption[justification=left]{Flux limiters and the 2nd-order TVD region. a) Plots of three flux limiters as they compare to the 2nd-order TVD region. Note that both the $8\times$-coarse-grained limiter learned with our machine learning methods and the van Albada 2 limiter extend outside of the 2nd-order TVD region. b) The evolution of a sinusoidal initial condition with first-order only (Lax-Friedrichs), second-order only (Lax-Wendroff), high-resolution first-order method (HR), and the three flux limiters plotted in a). Note that the $8\times$-coarse-grained limiter captures the shock more accurately than the others and, even though it does not lie in the 2nd-order TVD region, nonetheless exhibits no Gibbs effect (as does the Lax-Wendroff method shown in gray).}
    \label{fig:TVDRegion}
\end{figure*}


We chose to measure the quality of a discretized flux limiter by minimizing the mean-squared misfit for the limiter using 6 grid-point segments from the training data, averaged over all segments (as in Eq.~\eqref{eq:aveloss}). We showed that under this condition, when searching for a generally optimal limiter across a range of hyperparameters, flux limiters should be designed with the following rules-of-thumb: they should have 1) a fixed point at $(r,\phi) = (0,0)$, {\color{black}2) a first segment with a slope of $b_1 \sim CG$}, 3) a second segment with a negative slope $b_2 < 0.0$, followed by 4) a roughly linear region around {\color{black}$r = 1$, $\phi \in [0.5,1.0]$,} and $k \sim K/2$, and 5) a larger final region where $r_K \sim CG$.
{\color{black}It} is interesting to note that the $(r,\phi) = (1,1)$ constraint, which is due to a requirement of second-order accuracy of the shock capturing scheme and Lipshitz continuity of $\phi$ \cite{sweby1984high}, is not obeyed, yet these limiters, nonetheless, perform better than standard limiters.

A main advantage of our limiters seems to be the flexibility allowed by the multiple segments that make up their shape. As opposed to the bent or bowed shapes of standard limiters, multiple segment limiters allow a spring-like compression that, at least for Burgers' equation, allows for an improvement in their performance. 

In Fig.~\ref{fig:TVDRegion}a, we show how our machine learned $8\times$-coarse-grained limiter fits with respect to the 2nd-order TVD region \cite{sweby1984high}. Full containment within the 2nd-order TVD region is a sufficient, but not necessary, condition to eliminate the possibility of Gibbs effects in 2nd-order shock capturing schemes (see discussion around Eqs.~2.15 and 2.16 in \cite{sweby1984high}). To confirm numerically that our limiters do not exhibit a Gibbs effect, we plot the evolution of a sinusoidal initial condition for $8\times$-coarse-grained initial data. Note that the van Albada 2 limiter is partially outside of the 2nd-order TVD region in Fig.~\ref{fig:TVDRegion}a, and exhibits a distortion of the sinusoidal form in Fig.~\ref{fig:TVDRegion}b (particularly, away from the shock). Conversely, although our $8\times$-coarse-grained limiter is also not fully contained within the 2nd-order TVD region, it does not exhibit any visible Gibbs effect in Fig.~\ref{fig:TVDRegion}b and is entirely concave (resp. convex) on the left (resp. right) side of the shock. The $8\times$-coarse-grained limiter best captures the shock, and we see similar results for $2x$-, $3x$-, and $4x$-coarse-grained machine learned limiters (not shown).

Given the rules-of-thumb that we have observed, our results for finding an optimal limiter over the range of 
$CG \in [2, 10]$, {\color{black} $K \in [2, 38]$, and $\mu \in [0.005, 0.0248]$}, could most likely be improved by {\color{black}fixing the number of bins $K$,} 
constraining the limiter to follow the rules noted above, 
and rerunning using a number of simulations $N_s$ similar to that
used when solving for a fixed $(CG, K, \mu)$.

For fixed $(CG, K, \mu)$ calculations, we used $N_s = 500$ with $500$ random initial conditions, yielding in total 160M data points.
For variable $(CG, K, \mu)$ calculations, we used $N_s = 100$ with $100$ random initial conditions, using less than 30M data points.
We used a smaller $N_s$ for the hyperparameter calculations to help reduce the
computational cost of training the limiter, at the expense of some accuracy in
the learned flux limiter.
We performed our calculations on the Darwin cluster at Los Alamos National
Laboratory on a 22 core 
cluster, where simulations were run in parallel, one simulation per core. 

The first 12 iterations in the hyperparameter optimization completed in just less than 48 hours, with restarts (i.e. further optimization runs) performed over the next several days.
{\color{black}Fixing $K$, and
using a larger number of parallel cores, should be able to decrease the time to
obtain results.} {\color{black}
As the limiter segments that condense around $r = 1$ are found to be very short,
it}{\color{black} ~may be interesting to attempt to learn if there is
a relationship between the minimum number of bins required, the resolution
of the limiter, and the resulting standard error.} Another interesting study
may be to discover the impact on the observed set of rules when $N_c \ne 6$,
and if it has any impact on the shape or minimum number of bins.

Another interesting observation is that while we used a diffusion parameter
of $\nu = 0.01$ in our simulations, we found a learned diffusion parameter
of {\color{black}$\mu = 0.02471$} in our hyperparameter optimization (see Table \ref{tab:hyperparam}). {\color{black} The difference is }{\color{black}roughly}{\color{black} ~a factor of 2}, which counter-balances the difference in coarse-graining between our simulations and the learned limiter. Similar changes in effective diffusion are seen in analytical approaches to coarse-graining \cite{besnard1992turbulence,schwarzkopf2011application}. It may be prudent to determine if the diffusion parameter can always be determined similarly for other choices of CG.
\begin{figure*}[ht]%
    \vspace{-0.1 in}
    \centering
    \includegraphics[width=0.95\textwidth]{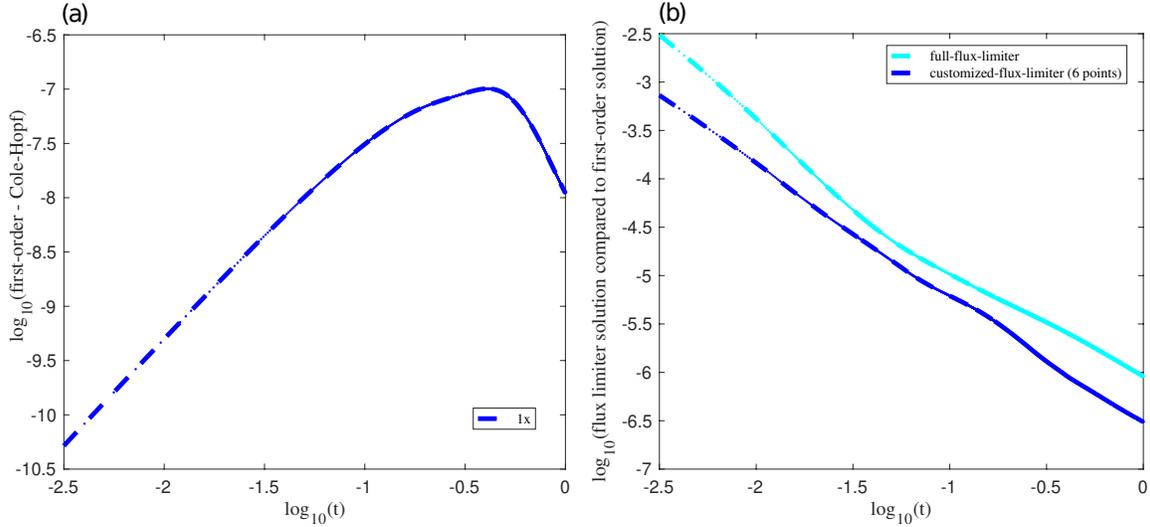}
    \vspace{-.0 in}
    \caption{a): Accuracy of the first-order, high-resolution solution as compared to the Cole-Hopf solution. See Sec.~\ref{apdx:first_and_exact}. b):  Relative error between the shock-captured solution computed with $\alpha =  \mathrm{max}(|\frac{\partial F}{\partial u})|_u$ and our customized Lax-Friedrichs flux with fixed $\alpha =  0.6$. See Sec.~\ref{apdx:customized_FL}
    }%
    \label{fig:first_and_exact}%
    \vspace{-.1 in}
\end{figure*}

Here, we chose to develop and test our machine learning approach to discovering improved flux limiters using $2$nd order shock-capturing methods. State-of-the-art methods such as the piecewise-parabolic methods \cite{colella1984piecewise,tripathi2017piecewise} (PPM, $4$th order) or (weighted) essentially non-oscillatory \cite{harten1997uniformly,liu1994weighted,ha2012mapped} (ENO, WENO, $8$th order) methods may also benefit from our machine learning approach.

Further, as the generation of training data contains some randomness, an improvement
would be to modify Eq.~\eqref{eq:loss} to train flux limiters that minimize the expected mean squared error between input-output pairs:
\begin{equation}
    C = \frac{1}{2} \sum_{i=1}^{N} { \mathbb{E}( o_i(\{ u_{c}^i \}) - g_i  )^2 } \; .
    \label{eq:expected_loss}
\end{equation}
In our study the impact of the randomness in the training data is mitigated by using a large number of simulations. Thus, we may be able to use significantly fewer simulations if we instead train the limiter on the expected mean-squared error.
Additionally, our approach could be used to train flux limiters for \emph{robustness} by minimizing $\overline{C}$, defined as:
\begin{equation}
    \overline{C} = \sum_{k=1}^{K} { \max C_{k} / K } \; ,
    \label{eq:robustness}
\end{equation}
for a range of $(CG, \mu)$, with $C_{k}$ defined as in Eq.~\eqref{eq:expected_loss}. 

\section{Conclusions}

We developed a theoretical framework that uses machine learning to train a continuous piecewise linear flux limiter by minimizing the mean-squared misfit to 6 grid-point segments of high-resolution data, averaged over all segments.
We demonstated our framework by producing a limiter that minimizes the misfit given a specific coarse graining, number of bins, and diffusion parameter.
We then compared the learned limiter to a set of 11 common flux limiters,
ranking the flux limiters by their log-error relative to high-resolution data.
We found that our $2 \times$, $K = 20$ bin machine learned limiter produces less misfit everywhere, with respect to all 11 other limiters.
Similarly, our $3 \times$, $4 \times$, and $8 \times$ limiters, with $K = 20$ bins, generally outperformed all the other flux limiters.
Our learned limiters all produced an initial kinked region with a large slope, a 
final kinked region with a smaller slope, and a smaller roughly linear region around $r_k \sim 1.0$ when $k \sim K/2$. Additionally, we see that generally $r_K \sim CG$ and $b_1 \sim CG$.
A main advantage of our limiters seems to be the flexibility allowed by the multiple line segments, as opposed to the bent or bowed shapes of standard limiters, enabling a spring-like compression that, at least for Burgers’ equation, yields an improvement in their performance.
We also extended our framework to demonstrate we can machine learn piecewise-linear flux limiters that outperform standard limiters across a range of hyperparameter values of coarse-graining, number of discretized bins, and diffusion parameter.
We found that learned limiters for the hyperparameter case also appear to {\color{black}generally} adhere to the same rules as the piecewise linear limiters trained at fixed parameter values.
Our study demonstrates a new approach to produce flux limiters that should be more broadly useful than standard limiters for general applications.

\section{Acknowledgements}
{\color{black}We thank Robert Chiodi for discovering and correcting an error in our numerical code, which led us to correct a portion of our study.} We thank Daniel Livescu for helpful discussions. Research presented in this article was supported by the NNSA's Advanced Simulation and Computing Beyond Moore's Law Program at Los Alamos National Laboratory, and by the Uncertainty Quantification Foundation under the Statistical Learning program. Los Alamos National Laboratory is operated by Triad National Security, LLC, for the National Nuclear Security Administration of the U.S. Department of Energy (Contract No. 89233218CNA000001). The Uncertainty Quantification Foundation is a nonprofit dedicated to the advancement of predictive science through research, education, and the development and dissemination of advanced technologies. This document's LANL designation is LA-UR-21-28444.

\appendix


%
\section{High-resolution solutions} \label{apdx:first_and_exact}
In this Appendix, we discuss the solution obtained with a high-resolution, first-order scheme that we use to generate training and testing data, as compared to the discretized Cole-Hopf approach.  We note again that the Cole-Hopf approach may exhibit instability when $\nu$ is small.  The explicit form of the high-dimensional solution to Eq.~\eqref{eq:burgers} was obtained numerically as:
\begin{equation}
    u_i (t_{n+1}) = u_i(t_n) + {\Delta t} ( \boldsymbol{D} \boldsymbol{v} + \boldsymbol{G} \boldsymbol{v}^2)
\end{equation}
with $\boldsymbol{D} = \frac{1}{(4 \Delta x)}  [1, 0, -1]$, $\boldsymbol{v} = [u_{j-1}, u_{j}, u_{j+1}]^T$, and $\boldsymbol G = \frac{\nu}{({\Delta x}^2)}  [1, -2, 1]$.  Here $\boldsymbol v^2$ indicates an element-by-element function giving the square of each element. Our high-resolution solution is very accurate when $\Delta t$ and $\Delta x$ stay small.  This is confirmed in Fig.~\ref{fig:first_and_exact}(a) where the high-resolution solutions are, at worst, within $10^{-6}$ of the discretized Cole-Hopf solution for coarse-grainings of $4 \times$, and yet better for coarse-grainings of $2 \times$ and below.  
Here, we use $\nu = 0.01$ and $\Delta t = 5 \times 10^{-4}$, $\Delta x = 5 \times 10^{-3}$ which are the values we used to generate our high-dimensional training dataset. 

\section{Customized Lax-Friedrichs flux}\label{apdx:customized_FL}
We explored methods to avoid computing the maximum over all $u_i$ when calculating the value of the partial derivative w.r.t. $u$ in Eq.~\eqref{eq:Lax-F}. If possible, this would allow us to reduce the training input points in our machine learning method.

We found that replacing $\alpha =  \mathrm{max}(|\frac{\partial F}{\partial u})|_u$ with the constant $\alpha =  0.6$ gave us improved solutions relative to the full flux-limiter when compared to our high-resolution data. 

In Fig.~\ref{fig:first_and_exact}(b) we plot the relative error between the custom flux limiter and high resolution data compared to the relative error between the full maximum flux limiter and high resolution data for a high resolution simulation with $\Delta x = 2.5 \times 10^{-3}$, $\Delta t = 2.5 \times 10^{-4}$, $\nu = 0.005$.
In the plot, we show the average solution error over twenty simulations with random initial conditions drawn from a uniform distribution from $v_\mathrm{min} = -1$ to $v_\mathrm{max} = 1$. Note that the modified method gives an error that is always less than the standard Lax-Friedrichs method.
Therefore, we use this customized flux, $f^\mathrm{low}$, with $N_c=6$ points, which appears in Eq.~\eqref{eq:o_i} in our machine learning model.

\section{Learned flux limiters }\label{apdx:full_coeffs}
Our discretized flux ratio bin edges, $\mathrm{r}_i$, and machine learned coefficients, ${b}_i$, for flux limiters corresponding to 2$\times$, 3$\times$, 4$\times$, 8$\times$ of coarse graining (see Fig.~\ref{fig:2x3x4x8x_FLfuncs_and_errors}(a)) are given in Tables~\ref{tab:20rs_2348x} and \ref{tab:20coeffs_2348x}. 
Note that bin width ($r_{i} - r_{i-1}$) decreases for small $r$ and increases for large $r$. Also note that $r_k \sim 1.0$ when $k \sim K/2$, $r_K \sim CG$, and $b_1 \sim CG$.
\begin{table*}[ht!]
{\footnotesize
\begin{tabular}{|c|c|c|c|c|c|c|c|c|c|c|c|c|c|c|c|c|c|c|c|l|l|}
\hline
   & $r_1$ & $r_2$   & $r_3$   & $r_4$   & $r_5$   & $r_6$   & $r_7$   & $r_8$   & $r_9$   & $r_{10}$  & $r_{11}$  & $r_{12}$  & $r_{13}$  & $r_{14}$  & $r_{15}$  & $r_{16}$  & $r_{17}$  & $r_{18}$  & $r_{19}$  & $r_{20}$  & $r_{21}$ \\ \hline
2x & 0  & 0.37 & 0.56 & 0.68 & 0.76 & 0.82 & 0.87 & 0.91 & 0.95 & 0.98 & 1.00  & 1.04 & 1.07 & 1.13 & 1.15 & 1.21 & 1.3  & 1.44 & 1.73 & 2.58 & 10  \\ \hline
3x & 0  & 0.29 & 0.46 & 0.59 & 0.68 & 0.76 & 0.82 & 0.88 & 0.92 & 0.97 & 1.00  & 1.05 & 1.10  & 1.15 & 1.22 & 1.31 & 1.44 & 1.65 & 2.08 & 3.34 & 10  \\ \hline
4x & 0  & 0.23 & 0.40  & 0.52 & 0.62 & 0.71 & 0.78 & 0.85 & 0.91 & 0.96 & 1.01 & 1.07 & 1.13 & 1.2  & 1.28 & 1.4  & 1.58 & 1.85 & 2.41 & 4.07 & 10  \\ \hline
8x & 0  & 0.14 & 0.26 & 0.38 & 0.48 & 0.58 & 0.67 & 0.76 & 0.85 & 0.94 & 1.02 & 1.12 & 1.22 & 1.35 & 1.51 & 1.73 & 2.06 & 2.59 & 3.64 & 6.79 & 10  \\ \hline
\end{tabular}
}
\caption{ Discretized space $\mathrm{r}$ for 2$\times$, 3$\times$, 4$\times$, and 8$\times$ of coarse graining. }
\label{tab:20rs_2348x}
\end{table*}

        

\begin{table*}[ht!]
{\footnotesize
\begin{tabular}{|c|c|c|c|c|c|c|c|c|c|c|c|c|c|c|c|c|c|c|c|c|}
\hline
   & $b_1$ & $b_2$  & $b_3$ & $b_4$  & $b_5$ & $b_6$  & $b_7$ & $b_8$  & $b_9$ & $b_{10}$ & $b_{11}$ & $b_{12}$ & $b_{13}$ & $b_{14}$ & $b_{15}$ & $b_{16}$ & $b_{17}$ & $b_{18}$ & $b_{19}$ & $b_{20}$ \\ \hline
2x & 2.33 & -1.68 & 1.17 & -0.22 & 0.36 & 0.03  & 0.16 & -0.02 & 0.52 & -0.23     & 0.28      & -0.15     & 0.03      & 0.09      & 0.14      & -0.04     & 0.20      & -0.08     & 0.29      & -0.09     \\ \hline
3x & 3.11 & -1.63 & 0.89 & -0.08 & 0.10 & 0.24  & 0.08 & 0.02  & 0.32 & -0.10     & 0.24      & 0.02      & 0.06      & -0.07     & 0.17      & -0.04     & 0.21      & -0.06     & 0.23      & -0.14     \\ \hline
4x & 4.02 & -1.81 & 0.67 & -0.04 & 0.19 & 0.25  & 0.12 & -0.05 & 0.37 & -0.02     & 0.12      & -0.01     & 0.11      & 0.16      & -0.01     & 0.10       & 0.14      & -0.04     & 0.23      & -0.16     \\ \hline
8x & 7.28 & -2.36 & 0.75 & -0.33 & 0.22 & -0.04 & 0.18 & 0.16  & 0.16 & -0.04     & 0.20      & 0.12      & 0.06      & 0.04      & 0.15      & 0.10      & 0.07      & -0.02     & 0.16      & -0.30     \\ \hline
\end{tabular}
}
\caption{Line segment slopes $\mathrm{b}$ obtained for K=20 with 2$\times$, 3$\times$, 4$\times$, and 8$\times$ of coarse graining.}
\label{tab:20coeffs_2348x}
\end{table*}




Similarly, {\color{black}solved} $r_i$ and $b_i$ in the optimization of a discretized machine learned limiter in the region defined by $CG \in [2, 10]$, {\color{black} $K \in [2, 38]$, and $\mu \in [0.005, 0.0248]$} (see Fig.~\ref{fig:hyperparam}(a)) are given in Tables~\ref{tab:hyperrs} and \ref{tab:hypercoeffs}. The corresponding values of $CG$, $K$, and $\mu$ are given in Table~\ref{tab:hyperparam}. 

\begin{table*}[ht!]
{\footnotesize
\begin{tabular}{|c|c|c|c|c|}
\hline
     & $CG$ & $K$ & $\mu$ & $\overline{C}$ \\ \hline
 0   & 4x & 20 & 0.02431 & 0.00002491 \\ \hline
 3  & 2x & 33 & 0.02366 &  0.00000547\\ \hline
 8  & 2x  & 34 & 0.02476 & 0.00000393 \\ \hline
 20  & 2x  & 38 & 0.02479 & 0.00000387 \\ \hline
 55 & 2x  & 37 & 0.02473 & 0.00000354 \\ \hline
 80 & 2x & 35 & 0.02474 & 0.00000352 \\ \hline
 110 & 2x & 36 & 0.02471 & 0.00000341 \\ \hline
\end{tabular}
}
\caption{ Coarse graining, $CG$, number of bins, $K$, diffusion parameter, $\mu$, and associated cost, $\overline{C}$, at selected iterations in the optimization of a discretized machine learned limiter in the region defined by $CG \in [2, 10]$, { \color{black} $K \in [2, 38]$}, and { \color{black}$\mu \in [0.005, 0.0248]$ }.  }
\label{tab:hyperparam}
\end{table*}
\begin{table*}
{\scriptsize
\centering
\begin{tabular}{|c|c|c|c|c|c|c|c|c|c|c|c|c|c|c|c|c|c|c|c|c|c|c|} 
\hline
    & $r_1$   & $r_2$    & $r_3$   & $r_4$   & $r_5$   & $r_6$   & $r_7$   & $r_8$   & $r_9$   & $r_{10}$  & $r_{11}$  & $r_{12}$  & $r_{13}$  & $r_{14}$   & $r_{15}$   & $r_{16}$   & $r_{17}$   & $r_{18}$  & $r_{19}$  & $r_{20}$  & $r_{21}$  & $r_{22}$   \\ 
\cline{2-23}
    & $r_{23}$   & $r_{24}$    & $r_{25}$   & $r_{26}$   & $r_{27}$   & $r_{28}$   & $r_{29}$   & $r_{30}$   & $r_{31}$   & $r_{32}$   & $r_{33}$   & $r_{34}$   & $r_{35}$   & $r_{36}$   & $r_{37}$    &     &     &      &      &      &      &       \\ 
\hline
110 & 0.0  & 0.35  & 0.52 & 0.63 & 0.71 & 0.76 & 0.80 & 0.83 & 0.86 & 0.88 & 0.90 & 0.91 & 0.93 & 0.94  & 0.96  & 0.97  & 0.98  & 0.99 & 1,00 & 1.01 & 1.02 & 1.03  \\ 
\cline{2-23}
    & 1.04 & 1.06  & 1.07 & 1.09 & 1.11 & 1.13 & 1.16 & 1.19 & 1.24 & 1.30 & 1.40 & 1.56 & 1.88 & 2.84  & 10.00 &  &       &      &      &      &      &      \\
    \hline
\end{tabular}
}
\caption{ {\color{black}Solved discretized bin edge locations, $r_i$}, in the optimization of a discretized machine learned limiter in the region defined by $CG \in [2, 10]$, {\color{black}$K \in [2, 38]$}, and {\color{black}$\mu \in [0.005, 0.0248]$}. The corresponding values of $CG$, $K$, and $\mu$ are found in Table \ref{tab:hyperparam}.}
\label{tab:hyperrs}
\end{table*}

\begin{table*}
\scriptsize{
\centering
\begin{tabular}{|c|c|c|c|c|c|c|c|c|c|c|c|c|c|c|c|c|c|c|c|c|c|} 
\hline
    & $b_1$    & $b_2$    & $b_3$    & $b_4$    & $b_5$    & $b_6$    & $b_7$    & $b_8$    & $b_9$    & $b_{10}$   & $b_{11}$   & $b_{12}$   & $b_{13}$   & $b_{14}$   & $b_{15}$   & $b_{16}$   & $b_{17}$   & $b_{18}$   & $b_{19}$   & $b_{20}$   & $b_{21}$  \\ 
\cline{2-22}
    & $b_{22}$     & $b_{23}$    & $b_{24}$    & $b_{25}$    & $b_{26}$    & $b_{27}$    & $b_{28}$    & $b_{29}$    & $b_{30}$    & $b_{31}$   & $b_{32}$   & $b_{33}$   & $b_{34}$   & $b_{35}$   & $b_{36}$   &    &    &       &       &       &    \\ 
\hline
110 & 2.11  & -1.58 & 1.10  & -0.27 & 0.49  & -0.16 & 0.51  & -0.05 & 0.09  & 0.37 & 0.75  & -1.11 & 0.36  & -0.37 & 0.98  & -0.33 & 0.04  & 0.75 & -1.37 & 1.27  & 0.77        \\ 
\cline{2-22}
     & -1.88   & 1.45 & -0.28  & 0.93  & 0.25  & 0.07 & 0.29  & -0.67 & 0.32  & 0.14 & 0.03 & 0.31  & 0.004  & 0.36 &  0.09 &  &       &       &       &       &      \\
\hline
\end{tabular}
}
\caption{ {\color{black}Solved line segment slopes, $b_i$}, in the optimization of a discretized machine learned limiter in the region defined by $CG \in [2, 10]$, {\color{black}$K \in [2, 38]$}, and {\color{black}$\mu \in [0.005, 0.0248]$}. The corresponding values of $CG$, $K$, and $\mu$ are found in Table \ref{tab:hyperparam}.}
\label{tab:hypercoeffs}
\end{table*}


\section{Standard flux limiters}

See Table~\ref{tab:fl_table} for the flux limiters used for comparison in Figs.~\ref{fig:FL} and \ref{fig:across_bin2x}.

\begin{table*}[ht!]
    \centering
    \begin{tabular}{r l}
        {\bf superbee} \cite{roe1986characteristic}: & $\phi_\mathrm{sb}(r) = \max(0, \min(2r, 1), \min(r,2))$ \\

{\bf monotonized central} \cite{van1977towards}: & $\phi_\mathrm{mc}(r) = \max(0, \min(2r, 0.5(1+r), 2))$ \\

{\bf smart} \cite{gaskell1988curvature}: & $\phi_\mathrm{sm}(r) = \max(0, \min(2r, (1/4 + 3r/4), 4)$ \\

{\bf Koren} \cite{koren1993robust}: & $\phi_\mathrm{kn}(r) = \max(0, \min(2r, \min((1/3 + 2r/3), 2)))$ \\

{\bf van Leer} \cite{van1974towards}: & $\phi_\mathrm{vL}(r) = \frac{r + |r|}{1 + |r|}$ \\

{\bf HCUS} \cite{waterson1995unified}: & $\phi_\mathrm{hc}(r) = \frac{1.5\;(r + |r|)}{r + 2}$ \\

{\bf ospre} \cite{waterson1995unified}: & $\phi_\mathrm{os}(r) = \frac{1.5\;(r^2 + r)}{r^2 + r + 1}$ \\

{\bf UMIST} \cite{lien1994upstream}: & $\phi_\mathrm{um}(r) = \max(0, \min(2r, 1/4 + 3r/4, 3/4 + r/4, 2))$ \\

{\bf van Albada 1} \cite{van1997comparative}: & $\phi_\mathrm{va1}(r) = \frac{r^2 + r}{r^2 + 1}$ \\

{\bf van Albada 2} \cite{kermani2003thermodynamically}: & $\phi_\mathrm{va2}(r) = \frac{2r}{r^2+1}$ \\

{\bf symmetric minmod} \cite{roe1986characteristic}: & $\phi_\mathrm{mm}(r) = \max(0,\min(1,r))$

    \end{tabular}
    \caption{Mathematical expressions for the standard flux limiters used for comparison in Figs.~\ref{fig:FL} and \ref{fig:across_bin2x}}
    \label{tab:fl_table}
\end{table*}

\bibliography{Biblio.bib}

\end{document}